\documentclass[a4paper,12pt]{article}
\usepackage{amssymb,amsfonts,amsmath,mathtext,cite,enumerate,float}
\usepackage{amssymb,amsthm,latexsym,euscript,amscd, authblk}
\usepackage{braket}

\usepackage[english]{babel}
\usepackage[cp1251]{inputenc}

\title {On linear structure of non-commutative operator graphs}

\author[1]{G.G.Amosov\thanks {gramos@mi-ras.ru}}

\author[1,2]{A.S. Mokeev\thanks {alexandrmokeev@yandex.ru}}

\affil[1]{Steklov Mathematical Institute of Russian Academy of Sciences, 8 Gubkina St., Moscow, 119991 Russia}

\affil[2]{Chebyshev Laboratory of St. Petersburg State University, 14th Line V.O. 29B, St. Petersburg 199178, Russia}

\begin{document}

\maketitle

\begin{abstract}
We continue the study of non-commutative operator graphs generated by resolutions of identity covariant with respect to unitary actions of the circle group and the Heisenber-Weyl group as well. It is shown that the graphs generated by the circle group has the system of unitary generators fulfilling permutations of basis vectors. For the graph generated by the Heisenberg-Weyl group the explicit formula for a dimension is given.
Thus, we found a new description of the linear structure for the operator graphs introduced in our previous works.
\end{abstract}

{\bf Keywords} non-commutative operator graphs, covariant resolution of identity, quantum anticliques

\section{Introduction} 

In this article we give a clarification of our current results on non-commutative operator graphs generated by covariant resolutions of identity \cite{amosovmokeev1,amosovmokeev2}. 
A linear subspace $\mathcal{V}$ in the algebra of all bounded linear operators on the Hilbert space $H$ is called a non-commutative operator graph if
$$
A\in\mathcal{V} \Rightarrow A^{*}\in \mathcal{V};\ I \in \mathcal{V}.
$$  
These objects allow to determine a possibility of zero-error correction for a system of unitary operators representing errors in quantum information transmission (for explicit description of this theory see \cite{Duan,amosovmokeev3}). Basically, we interested only in graphs satisfying the Knill-Laflamme condition \cite{lafl}. It is said that the graph $\mathcal{V}$ satisfies the Knill-Laflamme condition, if there is an orthogonal projection $P_K$ on the subspace $K\subset H$ such that $P_K\mathcal{V}P_K=\mathbb{C}P_K$, the projection $P_K$ is called an anticlique and the subspace $K$ is known as a quantum error correcting code.

Some properties and illustrating examples of non-commutative operators graphs could be given in terms of resolutions of identity covariant with respect to an action of locally-compact groups. Studying of non-commutative operator graphs generated by covariant resolution of identity was initiated in \cite{amo}.
 Let $G$ be a locally compact group and $\mathfrak{B}$ is a sigma-algebra generated by compact subsets of G, then the set of positive operators $\{M(B),B\in \mathfrak{B}\}$ is said to be a resolution of identity if
$$
M(\emptyset)=0,\ M(G)=I,
$$
$$
M(\cup_j B_j)=\sum_{j}M(B_j),\  B_k\cap B_l=\emptyset\  \textit{for}\  k\neq l,\  B_j\in \mathfrak{B}.	
$$
Let $(U_g)_{g\in G}$ be a projective unitary representation of $G$. Define the action of this group on operator $A$ as
$$
\rho_g(A)=U_gAU_g^{*}.
$$
Then $M(B)$ is said to be covariant with respect to this action if
$$
\rho_g(M(B))=M(gB).
$$
Suppose that a graph $\mathcal{V}$ is generated by $M(B)$ in the following sense
$$
\mathcal{V}=\overline{span\{M(B),B\in\mathfrak{B}\}}.
$$
Note that if $H$ is finite dimensional there is no need to take a closure in the last equation.

\section{Linear structure of $\mathcal{V}$}
At first let us define the graph introduced in \cite{amosovmokeev1}. Consider the finite dimensional Hilbert space $H={\mathbb C}^d\otimes {\mathbb C}^d$ and the orthonormal basis $\ket{k}, \ 1 \le k \le d$ in $\mathbb{C}^{d}$. 
Then, vectors
\begin{equation}\label{key}
\ket{\psi_ {sn}}=\frac{1}{\sqrt{d}} \sum_{k=1}^{d} e^{\frac{2\pi i s k}{d}}\ket{k\ k-n\  mod\  d}, 1\le s\le d,
\end{equation}
are known as the generalized Bell states.
Let $P_s, \ 1 \le s \le d$ be the projections on the subspaces
$$
H_s=span \{ \ket{\psi_{sn}}, \ 1 \le n \le d \}.
$$

Define a unitary representation of the circle group $\mathbb {T}=[0,2\pi]$ with the operation $+/mod(2\pi)$ by the formula
$$
\varphi \rightarrow U_{\varphi}= \sum_{s=1}^{d}e^{i \varphi s}P_s,\ \varphi \in \mathbb {T}.
$$

Also consider a set of projections in $\mathbb{C}^{d}$ of the form
\begin{equation}\label{Q}
Q_j=\sum \limits _{k=1}^d\ket {j\ j-k\ mod\ d}\bra {j\ j-k\ mod\ d},\ 1\le j\le d.
\end{equation}
The following corollary defines examples of the graphs mentioned above.

{\bf Corollary 1}\cite {amosovmokeev1}. {\it Given $j,\ 1\le j\le d,$ the projection (\ref {Q}) generates the graph $\mathcal{V}_j= span\{ U_\varphi Q_j U_\varphi^{*}, \varphi \in {\mathbb T} \} $ for which the projections $\{P_{s},\ 1 \le s \le d\}$ are anticliques.}

Let us consider a set of unitary operators defined as follows
\begin{equation}\label{W}
W_n=\sum \limits _{l=1}^{d}\sum \limits _{r=1}^{d}\ket {\psi _{l+n\ mod\ d\ r}}\bra {\psi _{lr}},\ 0\le n\le d-1.
\end{equation}
The unitaries from (\ref {W}) act as permutations on the set of vectors (\ref {key}).

Below we clarify a structure of the graphs determined by Corollary 1.

{\bf Theorem 1.} {\it All graphs $\mathcal{V}_{j}\equiv \mathcal {V}$ coincide for $1\le j\le d$. Moreover, one can choose
(\ref {W}) as generators of $\mathcal V$.}

{\bf Proof.}

For operators $U_{\varphi}Q_j U_{\varphi}^*$ we have such expansion in basis $\ket{\psi_{kn}}$ 
$$
U_{\varphi}Q_j U_{\varphi}^*=\sum_{s=1}^{d}\sum_{l=1}^{d}e^{i\varphi(s-l)}P_sQ_jP_l=
$$ 
$$
\sum_{s=1}^{d}\sum_{l=1}^{d}\sum_{r=1}^{d}e^{i\varphi(s-l)}P_s\ket{j\ j-r \ mod\  d}\bra{j\ j-r\  mod\  d}P_l=
$$
$$
\sum_{s=1}^{d}\sum_{l=1}^{d}\sum_{r=1}^{d}\sum_{n=1}^{d}\sum_{h=1}^{d}e^{i\varphi(s-l)}\ket{\psi_{sn}}\braket{\psi_{sn}|j\ j-r\  mod\  d}\braket{j\ j-r\  mod\  d|\psi_{lh}}\bra{\psi_{lh}}=
$$
$$
\frac{1}{d}\sum_{s=1}^{d}\sum_{l=1}^{d}\sum_{r=1}^{d}\sum_{n=1}^{d}\sum_{h=1}^{d}e^{i\varphi(s-l)}\delta_{nr}\delta_{hr}e^{-\frac{2\pi ij(s-l)}{d}}\ket{\psi_{sn}}\bra{\psi_{lh}}=
$$
\begin{equation}\label{result}
\frac{1}{d}\sum_{s=1}^{d}\sum_{l=1}^{d}e^{i\varphi(s-l)}e^{-\frac{2\pi ij(s-l)}{d}}\sum_{r=1}^{d}\ket{\psi_{sr}}\bra{\psi_{lr}}=\frac{1}{d}\sum_{n=0}^{d-1}e^{i\left(\varphi-\frac{2\pi j}{d}\right)n} W_{n},
\end{equation}
where unitaries $W_n$ are determined by (\ref {W}). To complete the proof we need to show that $W_n\in {\mathcal V}_j$ for all 
$j,\ 1\le j\le d$. Substituting $\varphi=\varphi _{kj}=\frac {2\pi {(k+j)}}{d}$ into (\ref {result}) we obtain
$$
\frac {1}{d}\sum \limits _{n=0}^{d-1}e^{i\frac {2\pi kn}{d}}W_n=U_{\varphi _{kj}}Q_jU_{\varphi _{kj}}^*\in {\mathcal V}_j
$$
for all $0\le k\le d-1$. The result follows. $\Box $

\section{Examples for Heisenberg-Weyl Group}

Now we want to give an analogous clarification of linear structure for the example of the non-commutative operator graph generated by the same technique implied for the Heisenberg-Weyl group. Unlike the circle group this group is non-commutative.   

At first we need to introduce the construction from \cite{amosovmokeev2}. Let $\mathfrak {H},\ dim\mathfrak {H}=d$ be a finite-dimensional Hilbert space with the fixed orthogonal basis $\ket {j},\ 0\le j\le d-1$. Unitary operators $S$ and $M$ on $\mathfrak{H}$
given by the formulae
$$
S\ket {j}=\ket {j+1\ mod\ d},\ M\ket {j}=e^{\frac {2\pi i}{d}j}\ket {j}
$$
define the representation of Heisenberg-Weyl group $G_d$ of rank $d$.

Consider $H=\mathfrak{H}\otimes\mathfrak{H}$ with the orthonormal basis $\ket{kj}, 0 \le k,j \le d-1 $. 
Also consider the orthogonal basis in $H$ consisting of entangled vectors
$$
	h_k^0=\frac{1}{\sqrt{n}}\sum_{j=0}^{d-1}e^{\frac {2\pi ikj}{d}} \ket{jj},
$$
$$
	h_k^j=(I\otimes S^{j} )h_k^{0}.
$$
Define subspaces in $H$ spanned by systems of the following entangled vectors
$$
H_j = span \{ h_k^j, 0 \le k \le d-1  \}.
$$
Let $\pi $ be the map transmitting $S$ and $U$ to unitary operators in $H$
as follows
$$
\pi (S)h_k^j=h_{k+1\ mod\ d}^{j},
$$
$$
\pi (M)h_k^j=e^{\frac{2\pi i }{d}k}h_k^{j}.
$$
It is shown in \cite {amosovmokeev2} that
the map $\pi $ can be extended to a unitary representation of the group $G_d$.

Also we need special sums of matrix units 
$$
y_{ml}=\sum \limits _{k=0}^{d-1}\ket {h_m^k}\bra {h_l^k}
$$
and their linear combinations
$$
h_0=\sum \limits _{m=0}^{d-1} y_{m m }\equiv I,
$$
$$
h_p=\sum \limits _{m=0}^{d-1} y_{m+p \ mod\ d \  m } + y_{m\ m+p \ mod\ d },\ 1\le p \le d-1.
$$
Resulting linear structure of the operator graph generated by representation $\pi$ was given in \cite{amosovmokeev2} as follows $\mathcal{V}=span\{h_p, 0\le p \le d-1\}$. Here we give the explicit value for a dimension of this graph.

{\bf Theorem 2.} {\it A dimension of the graph is given by the formula
\begin{equation}\label{dimm}
\begin{cases}
& \dim \mathcal{V}=\frac{d-1}{2}+1 \text{ for odd } d \\ 
& \dim \mathcal{V}=\frac{d}{2}+1\text{ for even } d 
\end{cases}
\end{equation}
}
{\bf Proof.}

Equation
$$
h_p=h_{p^{'}}
$$
or equivalently
$$
\sum \limits _{m=0}^{d-1} y_{m+p \ mod\ d \  m } + \sum \limits _{m=0}^{d-1}y_{m\ m+p \ mod\ d }=
\sum \limits_{m^{'}=0}^{d-1} y_{m^{'}+p^{'} \ mod\ d \  m^{'} } + \sum \limits_{m^{'}=0}^{d-1} y_{m^{'}\ m^{'}+p^{'} \ mod\ d }
$$
holds true if
$$
p'=d-p
$$
because a first summand in the left hand side equals to a second summand in the right hand side and vice-versa.
Thus, $\dim \mathcal{V}$ expresses by (\ref{dimm}). $\Box$

The next corollary follows from the proof of the theorem. 

{\bf Corollary 2.} {\it The graph $\mathcal{V}$ has the system of generators
\begin{equation*}
\begin{cases}
& \mathcal{V}=span\{h_p,\ 0\le p\le \frac{d-1}{2}+1\} \text{ for odd } d \\ 
&\mathcal{V}=span\{h_p,\ 0\le p\le \frac{d}{2}+1\}\text{ for even } d 
\end{cases}
\end{equation*}
}

\section{Conclusion}

We gave a clarification of the examples of the operator graphs given in \cite{amosovmokeev1}. It is shown that all these examples coincide and it is possible to choose unitary generators of a given graph, so we have the same situation as in \cite{amosovmokeev2}. For examples given in \cite{amosovmokeev2} we gave the explicit formula for dimensions of the graphs and specified their linear structure.

\section {Acknowledgments} The work is supported by Russian Science Foundation under the grant no. 19-11-00086.


\begin{thebibliography}{10}

\bibitem{amosovmokeev1}  
	 G.G.~ Amosov, A.S.~Mokeev, \textquotedblleft On non-commutative operator graphs generated by covariant resolutions of identity, \textquotedblright Quantum Information Processing \textbf {17} (12), 325 (2018). 
	
	\bibitem{amosovmokeev2}
	 G.G.~Amosov, A.S.~Mokeev,  \textquotedblleft On non-commutative operator graphs generated by reducible unitary representation of the Heisenberg-Weyl group, \textquotedblright International Journal of Theoretical Physics, doi: 10.1007/s10773-018-3963-4.

\bibitem{amosovmokeev3} 
	 G.G.~Amosov, A.S.~Mokeev, \textquotedblleft On construction of anticliques for noncommutative operator graphs, \textquotedblright Zap. Nauchn. Sem. S.-Peterburg. Otdel. Mat. Inst. Steklov. (POMI) \textbf {456}, 5--15 (2017); J. Math. Sci. \textbf {234} (3), 269-275 (2018).
	
	\bibitem{Duan} 
	 R.~Duan, S.~Severini, A.~Winter, \textquotedblleft Zero-error communication via quantum channels, noncommutative graphs and a quantum Lovasz theta function, \textquotedblright IEEE Trans. Inf. Theory. \textbf {59} 1164-1174 (2013).

\bibitem{lafl}
	 E.~Knill, R.~Laflamme, \textquotedblleft Theory of quantum error-correcting codes, \textquotedblright Phys. Rev. A \textbf {55} (2), 900 (1997).

	\bibitem{amo}
	 G.G.~Amosov, \textquotedblleft On general properties of non-commutative operator graphs, \textquotedblright Lobachevskii Journal of Mathematics \textbf {39} (3), 304-308 (2018).
	
	
	
	\bibitem{Weaver}
	\ N.~Weaver, \textquotedblleft A quantum Ramsey theorem for operator systems,\textquotedblright Proceedings of the American Mathematical Society \textbf {145} (11), 4595-4605 (2017).
	
			
		
		
	
\end{thebibliography}
\end{document}